\documentclass{aa}
\usepackage{natbib,amssymb}
\usepackage[pdftex]{color,graphics}
\usepackage{rotating,footmisc}
\bibpunct{(}{)}{;}{a}{}{,}
\newcommand{\solm}{M$_{\odot}$\ }

\title{First proper motions of thin dust filaments at the Galactic Center}

\author{K. Mu\v{z}i\'{c}$^{1,2}$, A. Eckart$^{1,2}$, R. Sch\"odel$^1$, L. Meyer$^1$, A. Zensus$^{2,1}$} 

\institute{1) I. Physikalisches Institut, Universit\"at zu K\"oln,
           Z\"ulpicher Str. 77,
           50937 K\"oln, Germany\\
           2) Max-Planck-Institut f\"ur Radioastronomie, 
           Auf dem H\"ugel 69, 
	   53121 Bonn, Germany\\
           \email{muzic@ph1.uni-koeln.de} }

\date{Received 18 August 2006  / Accepted 23 April 2007}
\begin{document}
\abstract
{L'-band (3.8 $\mu$m) images of the Galactic Center show a large number of 
thin filaments in the mini-spiral, located west of the mini-cavity and along the inner edge of
 the Northern Arm. One possible mechanism that could produce such structures is 
the interaction of a central wind with the mini-spiral. Additionally, we identify similar 
features that appear to be associated with stars.}
{We present the first proper motion measurements of the thin dust filaments observed in 
the central parsec around SgrA* and 
investigate possible mechanisms that could be responsible for the observed motions.}
{The observations have been carried out using the NACO adaptive optics system at the ESO VLT. 
The images have been transformed to a common coordinate system and features of interest were extracted. 
Then a cross-correlation technique could be performed in order to determine the offsets between the features with respect to
their position in the  reference epoch.}   
{We derive the proper motions of a number of filaments and  
2 cometary shaped dusty sources close (in projection) to SgrA*.
We show that the shape and the motion of the filaments does not agree with a purely 
Keplerian motion of the gas in the potential of the supermassive black hole at the position of SgrA*.
Therefore, additional mechanisms must be responsible for their formation and motion.
We argue that the properties of the filaments are probably related to an outflow from the
disk of young mass-losing stars around SgrA*. In part, the outflow may originate from 
the black hole itself. We also present some evidence and theoretical considerations that
the outflow may be collimated.}
{}

\authorrunning{K. Mu\v{z}i\'{c} et. al.} 
\titlerunning{Thin dust filaments at the Galactic Center}

\maketitle


\section{Introduction}\label{intro}

\begin{figure*}
\centering
 \resizebox{18cm}{!}{\includegraphics{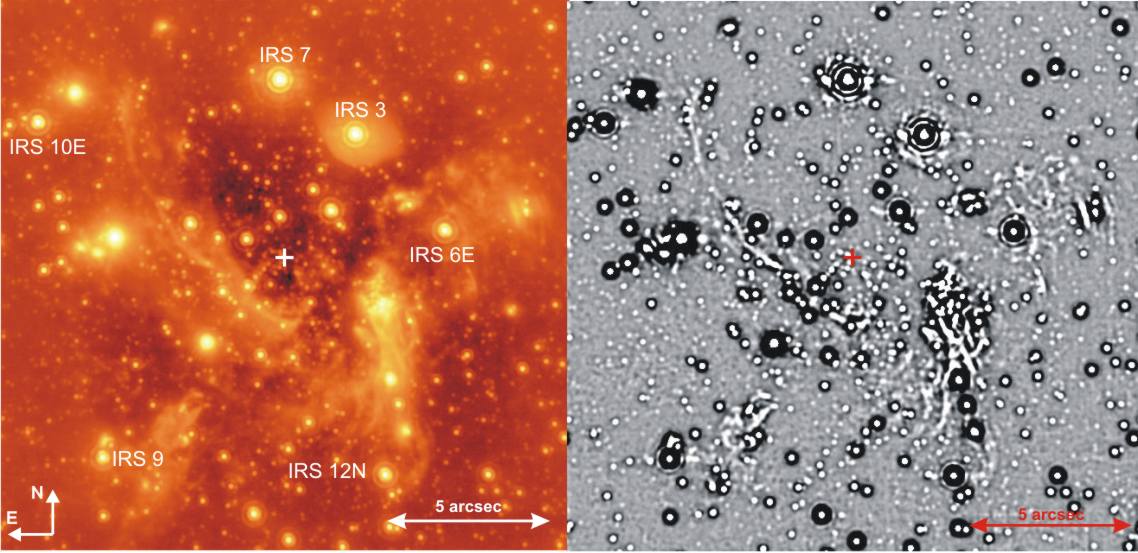}}
 \caption{Left: L'-band image of the central part of the Milky Way ($\sim$$\,$18$\,$$\times$$\,$18$\,$arcsec$^2$).  
The image contains a large number of dust embedded stellar sources, 
mini-spiral emission and several
previously unknown thin filaments. 
The image has a pixel scale of 27$\,$mas/pixel and resolution $\sim$$\,$100$\,$mas.
Right: High-pass filtered (smooth-subtracted) version of the image shown on the left. 
Here all compact features are enhanced. Cross in both images marks the position of SgrA*.}
\label{L-band}
\end{figure*}

At a distance of $\sim$ 8~kpc \citep{reid93,eisenhauer03,eisenhauer05}
the center of the Milky Way
shows a broad variety of structures.
Near-infrared diffraction limited imaging over the past 10 years
\citep{eckart96,genzel97,ghez98,genzel00,genzel03,ghez00,eckart02,schoedel03,ghez05}
has resulted in convincing evidence for a 3.6$\times$10$^6$\solm
black hole being located at the center of a dense stellar cluster.
This finding is supported by the discovery of a variable X-ray and NIR source
at the position of SgrA* \citep{bag01,bag03,eckart04}.
 
High angular resolution spectroscopic observations have confirmed 
the presence of a dense clumpy molecular ring (also called circum nuclear
disk, CND) of warm dust \citep{zylka95} and neutral gas 
\citep{guesten87,jackson93,wright01,hho02}
surrounding a short-lived central cavity of much lower mean gas density. 
The central cavity is about 3~pc in diameter \citep[see][]{rieke89,chan97,scoville03}
and contains 
the so-called mini-spiral, mainly composed of ionized gas and dust. \citet{vol&dus00}
 have re-examined the mini-spiral gas dynamics traced by the
[Ne$\,$II] line emission 
($\lambda$=12.8$\mu$m;  \citealt{lacy91})
and the H92$\alpha$ radio emission data \citep{rob&goss93}
and discuss the kinematic structure of the inner $\sim$~3$\times$4~pc 
of the Galaxy.  \citet{vol&dus00}
derive a three-dimensional kinematic model of gas streams that describes 
the bulk gas motion in three different planes.
Most of the material is located in a main plane which  
is connected to the inner edge of the CND.

\citet{paum04} observed the mini-spiral with integral field spectroscopy at 
2.06$\mu$m (He$\,$I) and 2.16$\mu$m (Br$\,$$\gamma$) covering roughly 40''$\times$ 40'' around SgrA*. 
They analyze the kinematics of the mini-spiral and describe nine distinguishable structures. 
From their analysis, the Northern Arm consists of a weak, continuous, 
triangular-shaped surface, drawn out into a narrow stream in the vicinity 
of SgrA* where it shows a strong velocity gradient, and a bright western rim. 

In the previously published observations of the mini-spiral, a number of narrow, 
filamentary structures 
 have been reported. The observations include high-resolution
radio data \citep{zhao&goss98,y-z98} and NIR 
Pa$\,$$\alpha$ \citep{scoville03}, 
Br$\,$$\gamma$ \citep{morris00} and He$\,$I \citep{paum01} emission line maps.
Here we present L'-band (3.8 $\mu$m; see Fig. \ref{L-band}) observations of the Galactic Center
 performed with the NACO adaptive optics system at the ESO VLT UT4
 \footnote{Based on observations collected at the European Southern Observatory, Chile}. 
 These data allow us not only to identify features as narrow as the $\sim$100~mas angular 
resolution of the observations and in much
larger number than in any previous dataset, but also to measure them clearly with high precision.
 Some of the filaments, particularly those 
associated with the Northern Arm and one detected stellar bow shock, have earlier been reported 
by \citet{clenet04} from observations at the same wavelength.

The L'-band flux from the mini-spiral is primarily coming from the 
thermal dust emission, which is confirmed by the observations at mid-infrared wavelengths \citep{viehmann06}.
 The detection of some of the filaments in Pa$\,$$\alpha$ and Br$\,$$\gamma$
tells us that the emission is as well associated with the ionized gas component of the ISM. 
In the L'-band, there is probably no significant contribution of the emission from the light scattered 
by dust grains. In this case one would expect to observe a significant
amount of the polarized emission also at the shorter wavelength K-band - which is not 
observed \citep{eckart95,ott99}. 
Furthermore, there is no bright NIR source at the Galactic Center that
could be identified as the source of large amounts of scattered light.
Polarized emission from the mini-spiral observed at longer infrared wavelengths is due
to thermal emission from elongated dust grains \citep{aitken91,aitken98}.

The central parsec of the Milky Way harbours a cluster of massive stars 
\citep{blum95,krabbe95,genzel96,genzel00,eckart99,clenet01,paum06} that supply
$\sim$$\,$3$\times$10$^{-3}$\solm$\,$yr$^{-1}$ of gas in the form
of stellar winds to the center \citep{najarro97}.
The amount of gas available for accretion onto SgrA* at the Bondi radius
is estimated to be $\lesssim$ 10$^{-5}$$\,$\solm$\,$yr$^{-1}$ \citep{quat03,bag03}.
The actual mass accreted onto SgrA* is up to 2 orders of magnitude lower, as determined
 by the linear polarization measurements at 230 GHz \citep{bower03, marrone06}.
We can see that about 99\% of the material from stellar winds 
does not even get close to the Bondi radius and must therefore
escape the central arcseconds in form of a wind. 
\citet{bland&begel99} point out that for accretion rates well below the Eddington rate most of
 the gas accretes with positive energy and may escape. Therefore they propose a solution 
(ADIOS; adiabatic inflow-outflow solution) in 
which most of the gas is lost through winds and only a small fraction of the gas 
is actually accreted onto the black hole.
 This is also
in agreement with the RIAF theory (radiatively inefficient accretion flow; see e.g. \citealt{yuan06}) 
that postulates strong outflows from SgrA*.
The outflow may be partially collimated in form of a jet.
Jet and shock models 
\citep[see][]{markoff03,melia01}
as well as accretion from an in-falling wind due to the surrounding cluster
of mass-losing hot stars \citep[e.g.][]{coker99,yuan02} are discussed
to explain the radio to X-ray properties of SgrA*.

There is some observational evidence for outflows in the central parsec as well.
The ''mini-cavity'' region 3.5'' southwest of the compact radio source SgrA*
was first pointed out on cm-radio maps by \citet{y-z90}.
The strong Fe[III] line emission seen toward that region 
\citep{eckart92,lutz93} is consistent with gas excited by 
a collision with a fast ($\ge$1000~km$\,$s$^{-1}$) wind from a
source within the central few arcseconds
\citep{y-z93,y-z92}. 
Most recently \citet{schoedel07} present an 
extinction map of the central parsec.
In this map an extended region of low extinction,
centered on SgrA* runs in NE-SW direction across SgrA* and 
is continued in the mini-cavity. It
supports the assumption of an outflow from the central arcseconds.
For one of the most prominent dust embedded sources, IRS~3, \citet{viehmann05} 
find that the extended L- and M-band continuum emission has a characteristic bow-shock shape. 
A possible explanation for its asymmetric appearance 
is that IRS~3 consists of a mass-losing star 
surrounded by a very thick, extended dust shell, 
which is pushed northwest by a wind
from the direction of the IRS~16 cluster and SgrA* 
\citep{pott07}.
Similarly to the north, the mass-losing envelope of IRS~7 appears to be
influenced by a strong central wind \citep{y-z92}.

The narrow filaments observed in the L'-band are the main concern of this paper: 
we identify them, present proper motion measurements, discuss their nature and 
present an interpretation 
in the light of the postulated outflow from the center.
In section \ref{obs} we present the observations. 
In section \ref{analysis} we identify thin filaments and explain
 proper motion measurements in detail.
In sections \ref{results} and \ref{discussion}  we describe the proper motions and discuss the  nature of 
the detected thin filaments.
In order to validate our proper motion results we compare the K-band and L'-band 
stellar proper motions in appendix \ref{stellar pm}.
In section \ref{wind} we present the probable wind sources in the central few arcseconds. 
In appendix \ref{outflow} we present the outflow model and discuss
the overall 3-dimensional orientation.
The summary and conclusions are given in section~\ref{summary}.

\section{Observations}
\label{obs}

 \begin{figure}
\centering
 \resizebox{9cm}{!}{\includegraphics{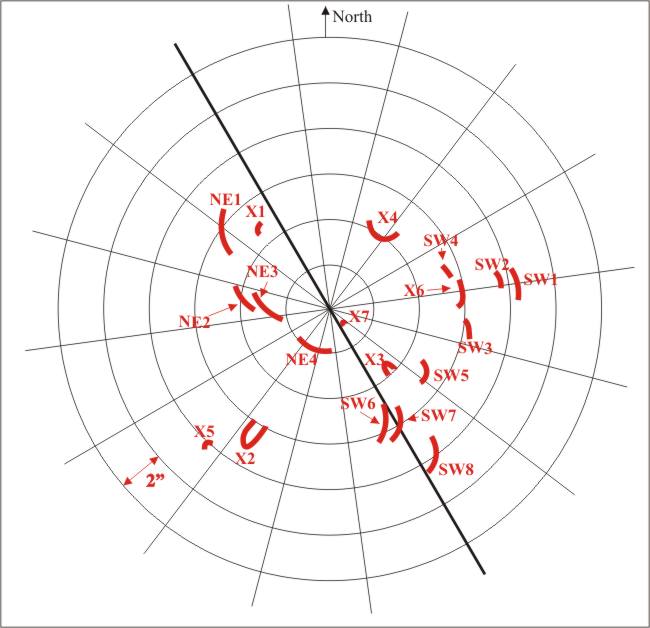}}
 \caption{Identification of the thin features observed in the the L'-band. SgrA* is located at the center of the image.
We have marked the line of nodes for disk $(i)$ at 
a position angle of 28$^o$ on the sky (N to E)
that contains the Northern Arm in the
model by \citet{vol&dus00}.
}
\label{ID}
\end{figure}

 \begin{figure}
\centering
 \resizebox{9cm}{!}{\includegraphics{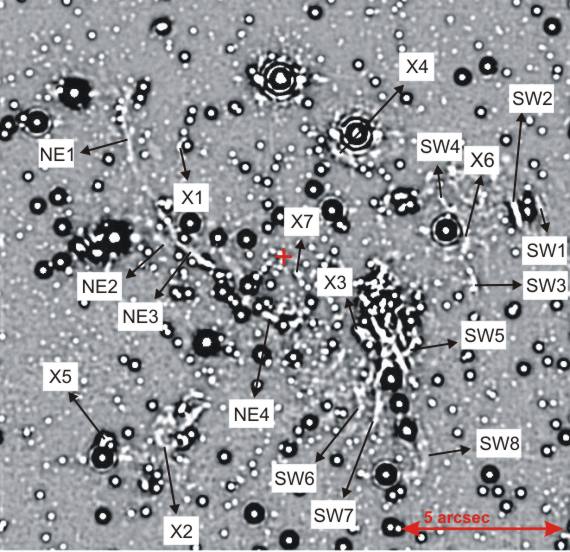}}
 \caption{High-pass filtered image with the narrow features labeled.}
\label{SSID}
\end{figure}
                    
The L' (3.8 $\mu$m) and K$_S$-band (2.1 $\mu$m)
images were taken with the NAOS/CONICA adaptive optics assisted imager/spectrometer 
\citep{lenzen98,rousset98,brandner02}
at the UT4 (Yepun) at the ESO VLT.
The L'-band data set includes images from 5 epochs 
(2002.660, 2003.356, 2004.320, 2005.364 and 2006.408) with a resolution of $\sim$100~mas. 
The K$_S$-band data set includes the images from epochs 2002.339, 2003.356, 2004.512, 2004.521 and 
2006.413 with a resolution of
 $\sim$56~mas.
Data reduction (bad pixel correction, sky subtraction, flat field correction) 
and formation of final mosaics was performed using the DPUSER software for 
astronomical image analysis (T. Ott; see also \citealt{eckart90}).  

The absolute positions of sources in our AO images were derived by comparison to the
VLA positions of IRS~10EE, 28, 9, 12N, 17, 7 and 15NE as given by \citet{reid93}.
The radio positions and the positions in the L'-band image agree to within 
less than a single 27~mas pixel i.e. a quarter of the beam. However, our 
proper motion measurements were based on relative astrometry between frames 
in order to avoid errors that may arise from uncertainties of the absolute positions.

\section{Data analysis}
\label{analysis}

\subsection{Identification of the thin filaments}

The L'-band images (Fig.~\ref{L-band}, left) clearly show stars, bright dust embedded 
sources as well as the emission of the mini-spiral.
In addition, one can distinguish a 
number of previously unknown thin filaments. 
In order to highlight these structures we produced high-pass filtered 
maps (Fig.~\ref{L-band}, right) by subtracting a smoothed version of the images from themselves, 
and then smoothing the final images. 
As a smoothing function in both cases we used a 4 pixel Gaussian (1 pixel corresponds
 to 0.027'').
This procedure enhances all structures that have significant 
power at spatial frequencies that correspond to the diffraction
limited beam size.
   
The thin filaments that are apparent in the L'-band maps are located
close to the mini-cavity, along the inner edge of the Northern Arm and, 
in some cases, in the vicinity of stars. 
The width of the filaments is $\lesssim$ 100~mas, the diffraction limit of NACO L'-band images.   
Fig.~\ref{ID} 
shows a schematic representation of the thin features in the central parsec of the Milky Way as
derived from the data  shown in the right panel of Fig.~\ref{L-band}.

According to their appearance we can divide the filaments in two classes: the first class 
represents features most probably associated with stars (denominated with X); the second class 
features are more elongated and associated with the mini-spiral. Those located west and south from 
SgrA* are denominated SW and those associated with the Northern Arm and located east 
from SgrA* we label NE. 
They cannot be associated with stars in a straightforward way.
We find that the second class of filaments to the east and northeast 
of SgrA* are located at the inner side of the Northern Arm  and curved with their 
convex sides eastward, whereas filaments to the west and southwest 
of SgrA* are curved with their convex sides westward and elongated almost perpendicularly to the Bar.
Several of the more luminous filaments to the west and southwest are
located on the western side of the mini-cavity close (in projection) to
the IRS13/IRS2 complex or just south of it (most prominent are SW5, SW6, SW7, SW8).
One can also notice that some of these filaments are located
up to about 5'' SSW of SgrA*, almost opposite to and approximately at the same 
distance as the most distant filament in the Northern Arm.
Concerning the X features, we notice no preferred direction towards which they are curved.
We discuss their possible nature in section~\ref{X}.

\subsection{Methods for proper motion measurements}
\label{methods}

In order to measure the positional offsets of a feature at different epochs, entire 
image frames had to be transformed to a common coordinate system. This was done using IDL 
image transformation routines. Transformation to second order was performed providing 
correction of all translations, rotations and possible distortions between two images. 
All the frames were transformed to the coordinate system of the 2003.356 epoch using 
a total of 34 stars to calculate the transformation matrix. The stars used for
 the transformation were chosen to be uniformly distributed across the field. The positions 
were corrected for the stellar proper motions as derived from the K$_S$-band images.
In order to validate our method and to demonstrate that reliable proper motions can be
obtained from the lower resolution L'-band data we compare K$_S$-band and L'-band stellar
proper motions in appendix \ref{stellar pm}.

In order to measure proper motions of the observed thin filaments we calculated offsets 
at all epochs with respect to the reference epoch (2003.356). First, we extracted an image 
section containing the entire feature, or 
 the part of the feature that is not directly contaminated by stars. 
 The position of the center and dimensions of the image sections
 are given in Table~1. Then we re-binned those image sections by a factor of 10 (since all the 
offsets are on sub-pixel scale), masked residual stars if present in the frame, 
and finally subtracted the background.
 The resulting small frames were cross-correlated with a reference frame in order 
to calculate the offset of the feature with respect to the reference position. 
We computed the inverse of the sum of the squares of the difference between 
overlapping pixel values. This resulted in a 2D cross-correlation function which was then fitted 
with a 2D Gaussian to determine the exact position of the peak that represents the best overlap of two features. 
There are two main sources of error that occur during the determination of the offset. 
The first is caused by the uncertainties due to the cross-correlation method. 
In order to determine this uncertainty for each feature and epoch, we shift it by different known sub-pixel 
offsets in all directions, run the cross-correlation routine and then calculate the 
standard deviation obtained from the imposed and derived shifts.
The second is resulting from the frame transformation procedure. 
From the initial sample of 34 stars we have randomly chosen 
50 different sub-samples of 20 stars each.
For all sub-samples we then repeated the transformation of all frames and calculated the positional 
uncertainty. 
The error bars in diagrams in Fig.~\ref{graphs} contain contributions of both sources of uncertainty.    
 
It is important to note that the extraction of
stars from our images via PSF fitting leaves significant residuals at the positions
of the brightest stars. Thus, our approach
was to extract sub-images containing the filaments from the original L'-band images if possible.
Due to the presence of a large number of stars in L'-band images, 
some of the identified features could not be extracted at all to perform 
the cross-correlation (e.~g. SW6, SW8, NE2). 
In some cases we were only able to extract one part of the filament, rather then the entire
structure. Besides the proximity of stellar sources, there is another reason for this: 
Many of the filaments are parts of more extended structures and 
could not be completely isolated.
This means that the positional data along the feature do not result in a measurable proper motion. 
Therefore, the velocities of all NE and SW features, except that of the compact features SW7 and NE4, 
are measured approximately perpendicular to their extent.
Some of the features are very faint in the L'-band and therefore 
unsuitable for the extraction. 
They can be more clearly identified in the smooth-subtracted images 
(features SW7, SW1, X4 and X6). 
In this case, only the SW7 appearance is sharp enough to give 
reliable results.  On the other hand, some of the features are 
very bright (X5 and SW2) and may be associated with embedded stellar sources.

Using the Starfinder code for stellar field analysis \citep{diolaiti00}
we also produced low-pass images containing the background emission. The result
 of the point source extraction largely depends on the size of the box (in units of a FWHM of the PSF) 
used for the background estimate. For small values of the box size ($<$ 10), the background estimate is still
contaminated by stars, and for values larger than 15 the compact features
become diffuse. The brightest of the Northern Arm filaments, NE4, has a high enough S/N to be clearly
detected in the background images.
The background image is free from stars and therefore allows us to
extract the NE4 feature without confusion. However, other NE and SW filaments cannot be
clearly defined against the smooth background. Thus, only for the NE4 filament we
additionally give the proper motion component along the extent of the feature (see also
appendix \ref{NE4}). We performed the NE4 proper
motion analysis for three different sizes of the box used for the background estimate (10, 13 and 15). 
The values given in Table~1 represent the average of the obtained results and the corresponding diagram
in Fig~\ref{graphs} shows only the result for the box with a width of 13 PSF FWHM values.

\section{Results}
\label{results}

 \begin{figure*}
\centering
 \resizebox{18cm}{!}{\includegraphics{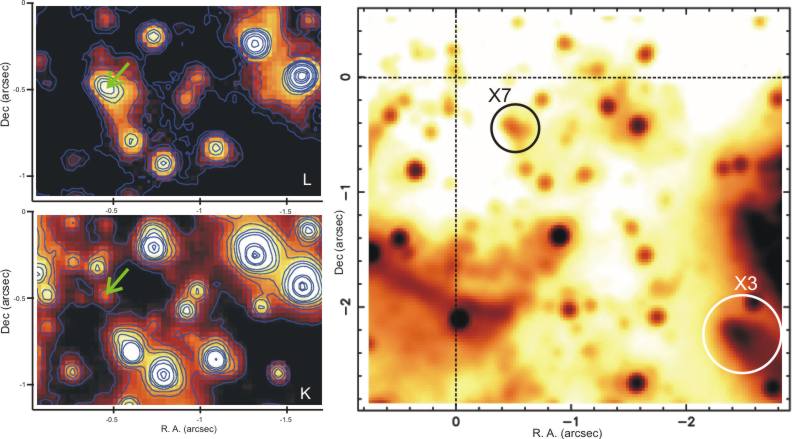}}
 \caption{Left: L'-band (up) and K$_S$-band (down) image of the small area 
southwest from SgrA* (axes show the offset from SgrA*).
Arrows mark the X7 feature in the L'-band and a star at the same 
position in the K$_S$-band frame. Right: L'-band image containing X3, X7 and SgrA*.}
\label{X3}
\end{figure*}

\subsection{Nature of the X features}
\label{X}
The features denominated with X are most probably associated with stars. 
The images clearly show that the features X1, X5, and X6 are curved, with stars at the 
approximate centers of their curvature - this suggests that they are associated with these stars.
X1, X5 and possibly X6 could be formed due to interaction
between a stellar wind, probably in combination with the stellar motion, 
and the interstellar medium. They could be similar to the Northern Arm stellar bow-shocks
discussed in \citet{tanner05}, but associated with less massive sources.
Feature X1 was already reported by \citet{clenet04}. The authors interpret
it as a bow-shock.
For feature X5 no significant proper motion was detected, but taking into
 account the significantly higher brightness of this feature with 
respect to the other filamentary structures, we suppose that there is 
an embedded or background star at the same position.   
The feature X4 is associated with IRS~3, whose bow-shock
appearance was discussed in \citet{viehmann05}.

Along the line that connects SgrA* and the feature X3 
($\sim$ 3.3'' southwest from SgrA*) at the position (-0.56'',~-0.55'') from SgrA*,
one can notice feature X7, with a shape similar to that of X3. 
The faint star at this position in the K$_S$-band image is represented
by a much more extended structure in the L'-band image (Fig.~\ref{X3}). 
Both X3 and X7 are approximately aligned (in projection) with SgrA*. 
The line of symmetry of the X7 ''V''-shape is oriented $\sim$40$^o$ east of north and to 
within less than 5$^o$ it is passing through the position of SgrA*.
The line of symmetry of X3 is slightly displaced from the line that connects it with SgrA* ($\sim$5$^o$) 
and is oriented  $\sim$37$^o$ east of north. The opening angle toward SgrA* is  $\sim$20$^o$ for X7 and  
$\sim$10$^o$ for X3. 
The orientation of both sources towards SgrA* suggests an interaction with a possibly existing 
outflow from the central region (see section \ref{probes}).
They may represent bow-shock-like features pointing back to the position of SgrA* as a common origin 
of such an outflow.

 \begin{figure*}
\centering
 \resizebox{18cm}{!}{\includegraphics{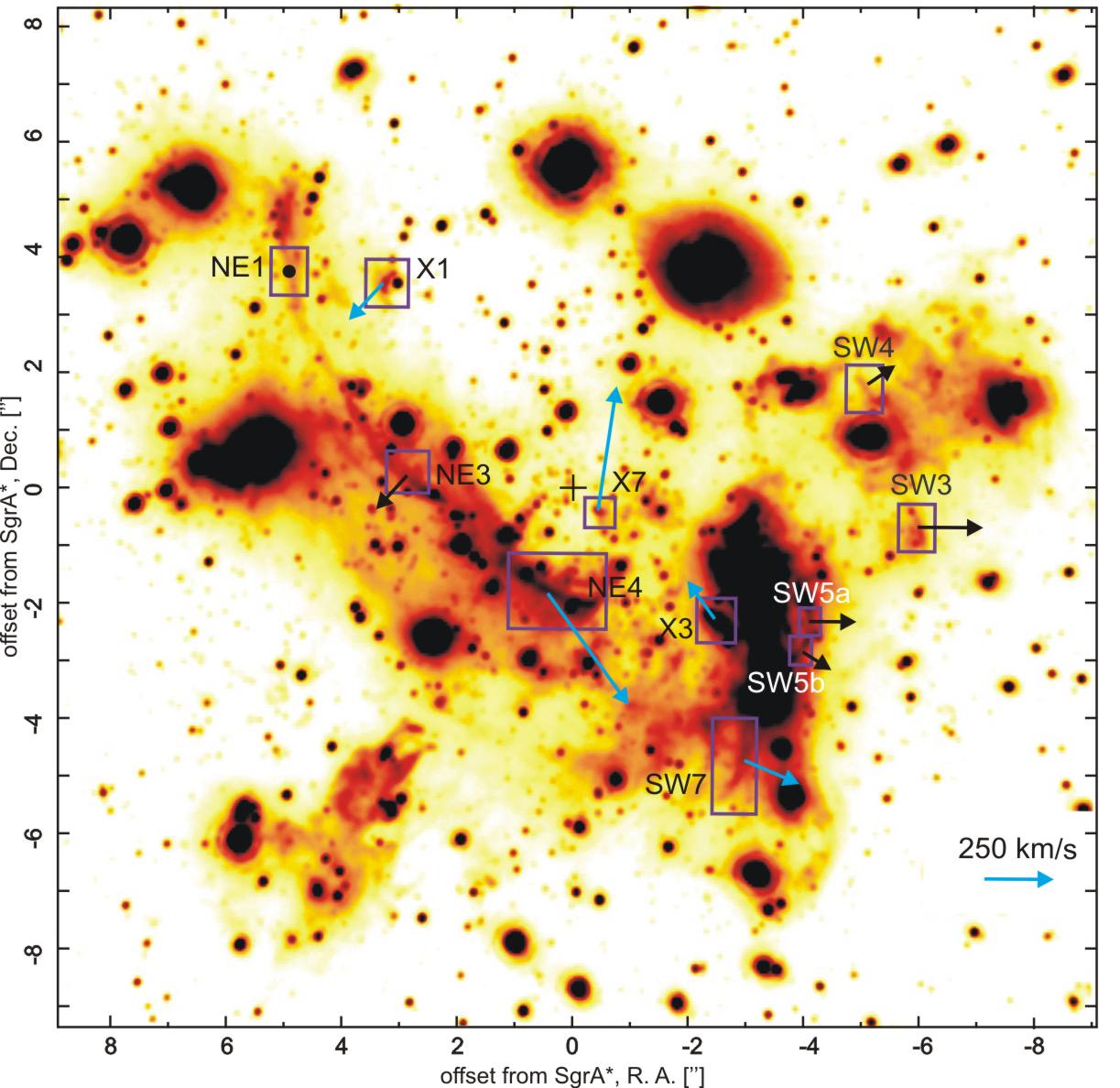}}
 \caption{L'-band image of the Galactic Center. Boxes mark thin filaments with measurable proper motions.
 Note that boxes in this image are different from those used for measurements (stated in Table~1.). The arrows show 
proper motions of the thin filaments obtained in our study: light blue arrows stand for the features with measurable 
proper motion in both directions, while black arrows show only the proper motion
component perpendicular to the feature
(see the text for the explanation). The insignificant motion 
of the Northern Arm filament NE1 is marked with a circle rather than an arrow. 
The cross marks the position of SgrA*.}

\label{proper motions}
\end{figure*}

\begin{table*}
\caption{Proper motions of thin filaments} 
\begin{center}
\begin{minipage}{1.0\textwidth}
\begin{tabular}{lccccccc}
\hline\hline
$feature$ & $\Delta\alpha$ \footnote{relative to SgrA*; position of the center of the box \label{fn3}}& $\Delta\delta$\footref{fn3} &
$size (\Delta x \times \Delta y)$& $v$ \footnote{velocities of all NE and SW features, except that of SW7 and NE4, are in a direction approximately perpendicular to the feature. Other proper motions are given both in R. A. and Dec. direction. In case of NE4 we additionally give the velocity along the feature and perpendicular to it. The positive sign in the velocity refers to the eastward motion in R.A. and northward in Dec. For the perpendicular component the positive sign marks the predominant eastward motion. The negative sign for the motion of the NE4 along the filament marks the predominant southward component. } & $\Delta v$ & $\phi$ \footnote {the position angle of the velocity vector for the features with a proper motion measured in the direction perpendicular to the feature. The angle is measured east of north.}&$v_{esc}$\\
 & (arcsec) &(arcsec) &(arcsec$\times$arcsec) &(km$\,$s$^{-1}$)&(km$\,$s$^{-1}$)& ($^{\circ}$) & (km$\,$s$^{-1}$)\\
\hline
NE1        &   4.76 &  3.42 & 0.32$\times$0.46 &  -29 & 16 &  90 & 368\\
NE3        &   2.62 & -0.09 & 0.38$\times$0.41 &  167 & 25 & 140 & 550\\
SW3       &  -6.08 & -0.77  & 0.35$\times$0.35 & -229 & 19 & 270 & 360 \\
SW4       &  -5.07 &  1.76  & 0.43$\times$0.38 &  -92 & 16 & 305 & 404\\
SW5a      &  -4.20 & -2.31  & 0.35$\times$0.11 & -165 & 15 & 270 & 407\\ 
SW5b      &  -4.13 & -2.75  & 0.16$\times$0.24 & -123 & 22 & 240 & 400\\ \hline 
NE4 (R.A.) &   0.50 & -1.63 & 2.51$\times$1.97 & -386 & 76 &  & 682\\
NE4 (Dec.) &        &       &                  & -280 & 55 &  &\\ 
NE4 (along) &       &       &                  & -436 & 103 & 240 & \\
NE4 (perp.) &       &       &                  &  194 & 11 & 150 &\\ \hline
SW7 (R.A.) &  -3.07 & -5.10 & 0.73$\times$1.22 & -186 & 17 &   &  365\\
SW7 (Dec.) &        &       &                  &  -80 & 21 &   & \\ \hline
X1  (R.A.) &   3.22 &  3.40 & 0.30$\times$0.38 &  113 & 16 &   & 412\\
X1  (Dec.) &        &       &                  &  -78 & 21 &   & \\ \hline
X7  (R.A.) &  -0.56 & -0.55 & 0.54$\times$0.41 &  -71 & 17 &   & 1006\\
X7  (Dec.) &        &       &                  &  458 & 21 &   & \\ \hline
X3  (R.A.) &  -2.45 & -2.37 & 0.46$\times$0.43 &  -78 & 18 &   & 483\\ 
X3  (Dec.) &        &       &                  &  135 & 20 &   & \\
\end{tabular}
\end{minipage}
\end{center}
\end{table*}
 
\subsection{Proper motion results} 
\label{proper motions res}

 \begin{figure*}
\centering
 \resizebox{18cm}{!}{\includegraphics{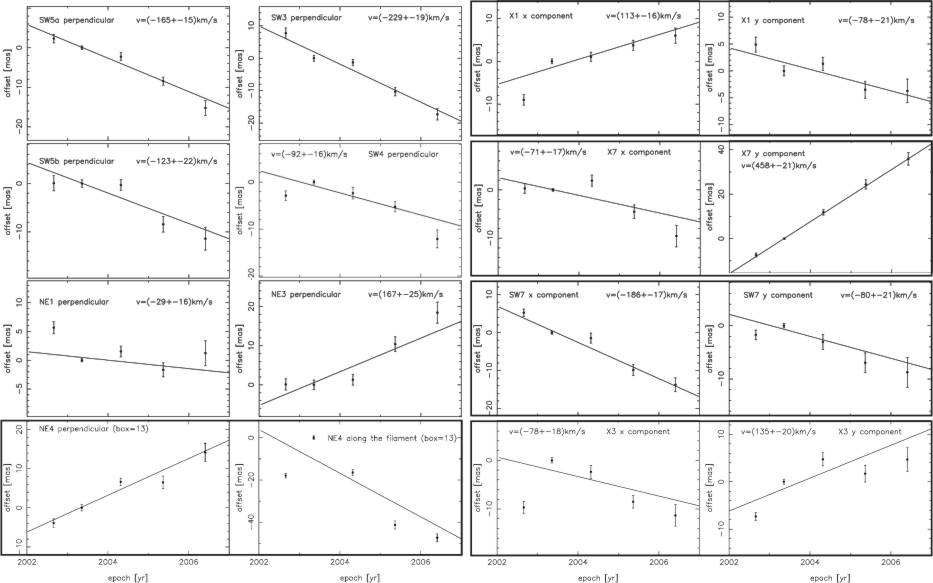}}
 \caption{Proper motions of thin filaments. The error bars show the 1 $\sigma$ uncertainty of each measurement.
The x and y components mentioned in graphs refer to R. A. and Dec., respectively.}
\label{graphs}
\end{figure*}

Table~1 summarizes the proper motions of thin filaments obtained in our study. 
The proper motions have been successfully determined in 11 regions. 
In order to transform the angular motions into velocities we assume a distance to SgrA* of 8~kpc. 

Boxes in  Fig.~\ref{proper motions} mark the filaments that were analyzed.  
The directions of motion in the plane of the sky are represented by arrows.
To emphasize that some of the motions shown in  Fig.~\ref{proper motions} do
not contain the complete information about the motion in the plane of the sky, we 
 represent them by different colors.
The light blue arrows in Fig. \ref{proper motions} show
 the full proper motion of the corresponding feature, while the black ones represent 
only the component perpendicular to the feature of interest. 
A filled circle rather than an arrow at the position of the NE1 feature marks the
insignificant motion of this feature. 
Note that for display purpose the dimensions of the boxes shown in  Fig.~\ref{proper motions} are 
different from those used for the extraction of the feature (as they are stated in Table~1).
Diagrams in Fig.~\ref{graphs} contain detailed information about the results.
The velocity difference $\Delta$$v$ in Table~1
represents the 1$\sigma$ uncertainty of the linear fit to the various epochs. Each offset is weighted by
the corresponding uncertainty. The escape velocities
were determined using the projected distance from SgrA* with a mass of 3.6$\times$10$^6$\solm.
  
Feature X1 is associated with a star located (2.95'',~3.46'') from SgrA*. 
The proper motion of the associated star is v$_{R.A.}$=(247$\pm$25)$\,$km$\,$s$^{-1}$, 
v$_{Dec.}$=(-46$\pm$5)$\,$km$\,$s$^{-1}$ 
as derived from our L'-band data.
  We find that the feature is moving approximately in the same direction as the star apparently 
associated with it - which confirmes the bow-shock interpretation of \citet{clenet04}.

\subsection{Comparison to Radio Measurements}
\label{radio}

\citet{y-z98} presented radio continuum observations at $\lambda$ =2~cm and calculated
 proper motions of ionized gas at the Galactic Center.  Although no thin filaments similar to those 
detected in NACO L'-band images are detected in the radio continuum images, some comparison can be done. 
The filamentary ionized structure connecting IRS~13 and IRS~2 (Box~11 in \citealt{y-z98})
 coincides with the SW5 feature in our images (in this work divided in two parts, SW5a and SW5b,
 because of the presence of a star). A westward direction of motion is obtained from 
both measurements, but the value of this component obtained from the radio continuum is larger 
(v$_{R.A.}$=(-329$\pm$56)$\,$km$\,$s$^{-1}$). The reason for this discrepancy is not clear. It is possible
that at different wavelengths we are actually probing different material.
\citet{zhao&goss98} also performed radio continuum observations at 7 and 13~mm and calculated
 proper motions of 57 compact H~II components. As in the previous case, the only comparable 
structure is SW5 (''loop'' in their notation), but the results are not in good agreement with ours. 
This feature, in the measurements of \citet{zhao&goss98}, 
is divided in several non-consistently moving components, 
which is probably consequence of the granular substructure 
seen in the radio image
(probably due to a low signal to noise ratio). 
In our IR continuum images the structure is well defined.

\citet{y-z98} report an anomalous high-velocity feature, so-called ''bullet'', 
at the offset (-3.39'', 2.065'') from SgrA* in the 1990 epoch. Approximately at this position
in the L'-band images one can notice a dusty feature (see Fig. \ref{proper motions}). However, 
 given the high velocity quoted for the bullet it should have moved by 
($\Delta$$\alpha$, $\Delta$$\delta$)$\,$=$\,$(0.30'',~0.34'') to the northwest 
from 1990 to 2006, which is the epoch of the image
 in Fig. \ref{proper motions}. No similar feature can be found at the corresponding
 position in the 2006 epoch, or close to it. 
Although the feature we see in the L'-band cannot be disentangled from the rest of the material
and therefore we are not able to measure its proper motion, a simple inspection
of images at different epochs shows that it does not seem to have a significant 
velocity compared to the bullet. Therefore, we cannot identify this feature as the bullet.

\section{Discussion}   
\label{discussion}

\subsection{Orbits of the diffuse gas}

\begin{figure}
\centering
\resizebox{9cm}{!}{\includegraphics{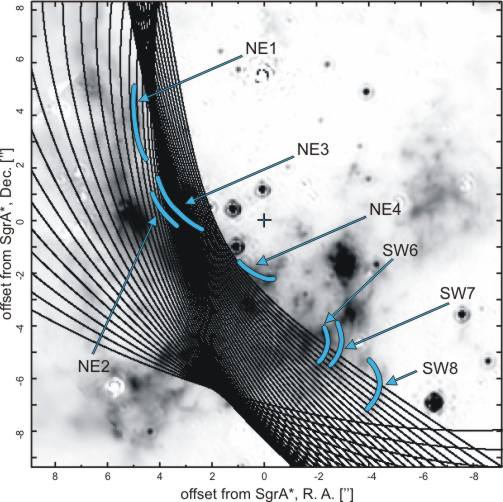}}
\caption{The Pa$\,$$\alpha$ map with the Keplerian orbits overplotted \citep{paum04}. We
additionally sketch some of the L'-band narrow filaments, showing that none of the orbits is aligned with 
the Northern Arm filaments.}
\label{orbits}
\end{figure}

The overall kinematic properties of the Northern Arm and Bar section 
can be modeled as a single gas streamer orbiting SgrA*.
Such a model can be constrained via the radial velocity information from 
the H92$\alpha$ \citep{rob&goss93} and [Ne$\,$II] line 
\citep{lacy91,vol&dus00} as well as the overall distribution of the gas 
and dust along these features.
Proper motions of ionized gas were in detail studied by \citet{y-z98}. 
They additionally give the radial velocities as measured by \citet{rob&goss93} and 
\citet{roberts96}. According to \citet{y-z98} the predominant component 
of the motion in the plane of the sky is from east to west. The flow of gas in the Northern Arm
follows its elongated shape: from the north-east with slightly redshifted 
velocities it follows the orbital trajectory to
the southwest as it crosses the plane of the sky. 
The ionized gas passes south of SgrA* before it moves to the northwest.
The complex velocity structure of the
mini-cavity, however, fails to be explained by such a simple model.
Similar results are reported by \citet{paum04} based on He$\,$I and Br$\,$$\gamma$
observations. At the position of the NE1 filament the line emission is
slightly red shifted (40-60 kms$^{-1}$) while at the positions of the other 
Northern Arm filaments the radial velocity becomes negative. At the position
of NE3 the line emission is blue shifted by 40-60 kms$^{-1}$ and close to NE4 
 by more than 130 kms$^{-1}$.
The Bar material at the position of SgrA* and within few arcseconds south of it 
has radial velocities close
to 0 kms$^{-1}$. The emission becomes blue shifted as we move further to the west 
(where SW3 and SW4 are located). 

\citet{paum04} analyzed the Northern Arm as a Keplerian system in the gravitational
field of SgrA*. They fit a bundle of Keplerian orbits in order
to model the full velocity field of the Northern Arm. 
The authors interpret the rim of the Northern Arm in terms of 
line-of-sight orbit crowding. The western edge facing SgrA* is
the region that shows the highest density of crowded orbits 
(see Fig. 8 in \citealt{paum04}). In order to construct Fig. \ref{orbits}
we used the part of the Fig. 8 in \citet{paum04} having the same field of view
 as the L'-band images presented in this work.

\subsection{Stability of the filamentary structures}
\label{stability}

The nature of the observed thin features, especially the SW and NE filaments, is unclear.
One possible interpretation is that they are the consequence of the interaction
 of a central wind originating from the hot stars and/or possibly from SgrA*. 
While the origin and the mechanisms of such a wind will be discussed later in the text, 
here we concentrate on the effects it would have on the mini-spiral material
in combination with the dominating gravitational potential of the central black hole.

We favor a scenario in which
the proposed wind compresses gas and dust in the mini-spiral and forms the observed filaments.
For the SW filaments this interaction could have taken place at the western edge of the 
mini-cavity. The thin Northern Arm filaments may have been generated by a wind in the 
opposite direction of the mini-cavity.
For the mini-spiral \citet{vol&dus01} quote a sound speed of v$_s$=7.6$\,$km$\,$s$^{-1}$. 
This velocity is very small in comparison with the overall motion of 
the Northern Arm \citep{paum04,y-z98}.
Also, once formed, the filaments may still be 
 influenced by wind from the central few arcseconds 
that may keep them confined.
The magnetic field of the Northern Arm, as reported by \citet{aitken91} and
\citet{aitken98},
could also play a role in confining ionized gas and charged dust grains along the 
filaments. 

If we assume a temperature of T=7000K for the mini-spiral material 
\citep{rob&goss93} then the average velocity of a hydrogen atom associated 
with thermal energy is v=(3kT/m)$^{1/2}$ $\approx$ 13$\,$km$\,$s$^{-1}$.  
At some position in the Northern Arm it takes approximately
250 yr for it to cross the distance equivalent to the width of a filament
that we can resolve ($\sim$100$\,$mas).
During this time the filament can drift $\sim$2.5'' soutwards due to the motion of
the mini-spiral material in the central gravitational potential. 
For dust particles it takes even longer to cross the
equivalent width of a filament due to their thermal motion.
This leads to the conclusion that a filament formed in the Northern Arm 
of the mini-spiral will have a lifetime that is long enough to drift to the 
southwest together with the gas stream.
Once formed, the dominant source for their 
destruction will be the interaction with their surrounding turbulent 
medium. Judged from the line widths of the H92$\alpha$ emission 
\citep{rob&goss93} the velocity dispersion of the gas within the 
mini-spiral is of the order of 50 km$\,$s$^{-1}$. 
\citet{vol&dus00} assume a maximum value for the velocity 
dispersion of 40\% of the Keplerian velocity.

\begin{figure}
\centering
\resizebox{9cm}{!}{\includegraphics{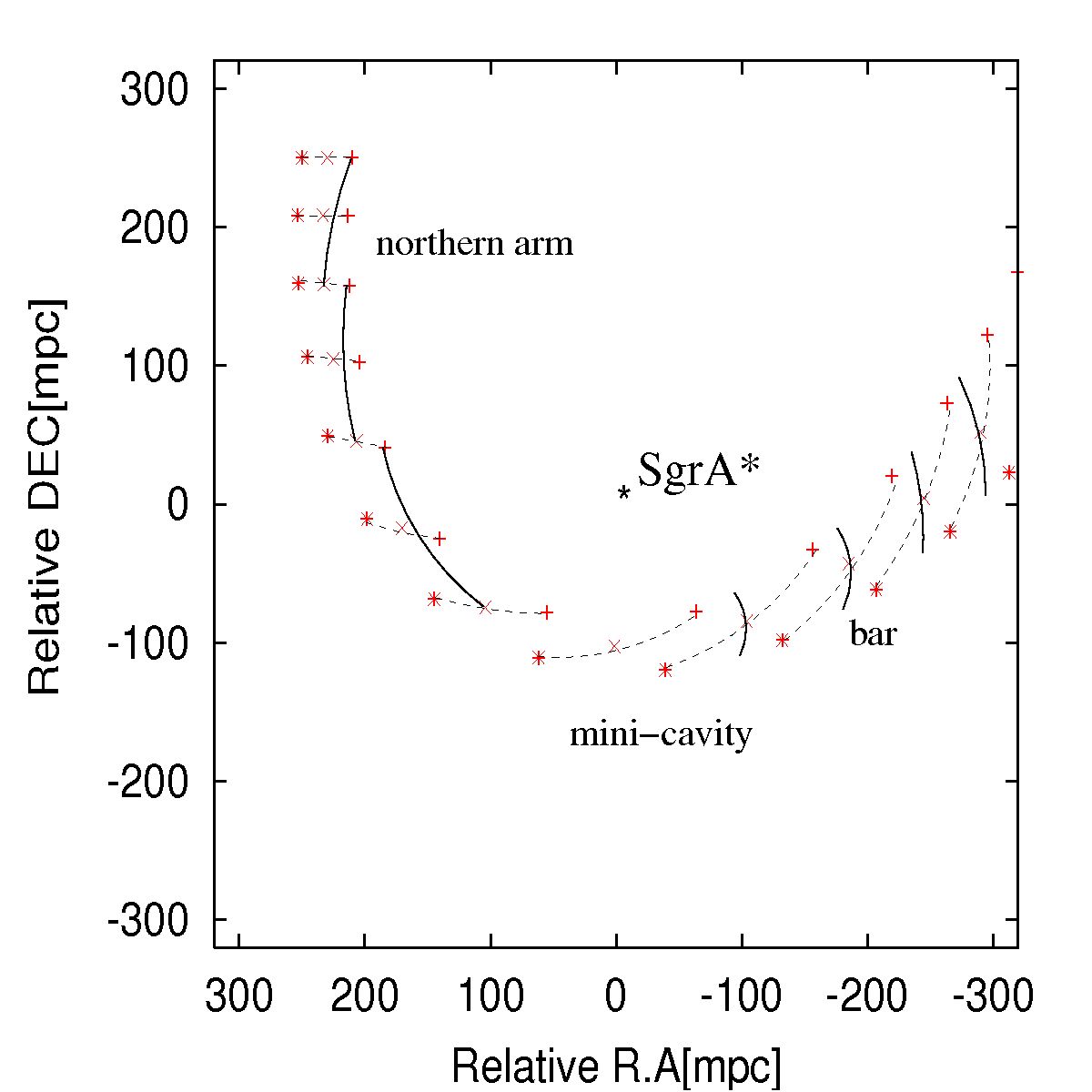}}
\caption{Results of our preliminary 3D dynamical model calculation. 
Three particles (marked +, x and *) have been placed within 
the mini-spiral with locations and velocities consistent 
with our current knowledge. See text for further explanation.
The dashed lines connect particles launched at equal times.
The time beetween the sets of symbols is $\sim$250$\,$yr.
The solid lines represent 
the observed narrow features (i.e. shock fronts or contact surfaces)
and how they evolve while travelling along the Northern Arm/Bar region.
A filament formed at some position in the Northern Arm or Bar 
is slightly changing the shape while drifting with the rest of the 
material, but preserves its filamentary appearance.
}
\label{drift}
\end{figure}

In order to investigate the kinematic behaviour of
linear features formed in a central potential
we carried out orbit simulations.
We assumed a dominating central potential
from a 3.6$\times$10$^6$\solm black hole
at the position of SgrA*.
\citet{schoedel07} estimate
that the gravitational field at the Galactic Center is dominated by the point mass of
the black hole inside a projected radius of about 6''. Since all of the measured filaments 
are either inside or just at the border of this circle, we do not take into account 
the possible contribution of the
stellar cluster to the gravitational potential.
In agreement to similar calculations by
\citet{vol&dus00}
and
\citet{paum04}
we chose 3 test particles with velocities of
$v_{R.A.}$=$\,$20$\,$km$\,$s$^{-1}$,
$v_{Dec.}$$\,$=$\,$-150$\,$km$\,$s$^{-1}$ and
$v_{LOS}$=$\,$-100$\,$km$\,$s$^{-1}$
launched from locations around
$\alpha_{relative}$ =$\,$220$\,$mpc$\,$$\,$=5.6'',
$\delta_{relative}$ =$\,$250$\,$mpc$\,$=$\,$6.4'' and
$z_{LOS}$$\,$=$\,$150$\,$mpc 
(see Fig.~\ref{drift}).
Here $z$ is the distance behind the plane of the sky including SgrA*.
The corresponding orbits show that 
elongated thin features ''imprinted'' into the Northern Arm
are subject to a weak shear but to first
order keep their original structure while drifting to the south.
In our current picture Northern Arm
filaments drift south (while being stretched by less than a
factor of 2 within the Northern Arm)
until they reach the mini-cavity region where
they may be destroyed. Filaments created at the western edge of
the mini-spiral drift westward along the Bar.

From Fig. \ref{orbits} one can clearly see that the Northern Arm filaments 
cannot be simply explained as the consequence of orbit-crowding as suggested by \citet{paum04}. 
A comparison with the velocity field obtained from their model shows a good agreement for the 
perpendicular velocity component of the filaments NE1 and NE3 (within a 1$\sigma$ error for NE1 and 
2$\sigma$ for NE3). This supports the assumption that the observed filaments are really 
drifting southwards 
together with the Northern Arm material. 
However, none of the orbits actually follows the shape of the filaments and therefore
the proper motions of the gas as stated in \citet{paum04} are not aligned with the extent 
of the filaments. Thus one needs to include an additional 
force that could form the filaments, keep them confined 
and prevent them from being destroyed by the bulk gas flow. 
This is especially valid for the filament NE3.
The exact position of the filament NE4 does not coincide with any of the Keplerian orbits 
fitted by \citet{paum04}. They find that the Northern Arm is split into 
two layers as it reaches the mini-cavity. The mini-cavity itself is a part of the first layer, 
while the second layer is deflected northward from it and seems to contain the filament NE4.   
Concerning the SW filaments, our study shows that the proper motion components perpendicular to their extent 
are consistent with the motion in the dominating central potential,
 but we are still facing the question of their formation. 
As proposed earlier in this section, they could have been formed by an interaction at the western 
edge of the mini-cavity, and then drift 
westwards together with the rest of the mini-spiral material.

Concerning the possibility that the filaments are the consequence of an
interaction between a fast wind and the mini-spiral material, we can try to
 predict the width of a shock using formula (8) in \citet{hartigan87}.
Assuming shock velocities 100-200$\,$km$\,$s$^{-1}$ and a density of the pre-shock medium $\sim$$\,$10$^3$cm$^{-3}$,
 we get the cooling distance of the shock to be $\sim$$\,$30$\,$AU$\,$$\sim$$\,$4$\,$mas, much smaller then
the resolution of the current observations.

Since the motions of the Northern Arm filaments fail to be completely 
explained simply by gravitational influence,
 one has to include additional forces that could drive the observed motions. 
 In the next section we consider the winds at the Galactic Center as the 
main candidates.

\section{Winds at the Galactic Center}
\label{wind}

As already discussed in the introduction, less than 1\% of the material supplied by massive stars
in the central parsec of the Galaxy is actually available for the accretion onto SgrA*. Most of the material  
 never gets close to the BH, i.e. it must be blown away from 
the center and therefore can interact with the material of the mini-spiral.

\subsection{Evidence for a central outflow}
\label{evid}

There is a body of observational evidence that supports the interaction of a wind
with the local ISM as well as with stars in the central parsec. 
Some of those were already presented in the introduction 
(mini-cavity, IRS~3, IRS~7, low extinction region).
Then, the features presented in this work: 
the narrow filaments could be the contact surfaces between the wind and the 
mini-spiral material. The ''V''-shaped sources X3 and X7 are supporting the SgrA* wind scenario
(see section \ref{probes}).

The main candidates for the source of the central outflow are the high-mass-losing stars
at the center. Alternatively, the outflow could be due to a jet-like structure directly
associated with the black hole at the position of SgrA*. Some of the theoretical models
that successfully fit the observational data include such a jet
 \citep[e.g.][]{falcke&markoff00,yuan02}.
In appendix \ref{outflow} we consider the possibility that the outflow could, in fact, 
be partially collimated.

\subsubsection{Stars as the source of the outflow}
\label{stars}

Most of the above mentioned bright and massive He-stars are located 
within a single plane and most of them within a radius of 0.1~pc (2.5'') from SgrA* 
\citep{genzel00,genzel03,l&b03}, 
forming a disk of clockwise rotating stars (CWS).
\citet{paum06} largely expand the sample of stars 
belonging to the CWS disk, and also propose 
the existence of a second, less populated stellar disk containing counter-clockwise rotating 
stars (CCWS; see also \citealt{genzel03}). 
According to \citet{paum06} most of the bright stars in the IRS~16 complex are part of 
the clockwise system, including IRS~16C and IRS~16SW. 
Since the bright He-stars in the CWS are also those that show clear indications for P~Cygni profiles 
we assume that the stars in this disk are also the predominant source of mass loss in the central parsec.
Since the geometry of the CWS plane is such that the rotational axis
is inclined towards the observer, the bright He-stars to the north and
 northeast of SgrA* are located on the far 
side of the  disk (i.e. IRS~16C, IRS~16CC, E29, E38; \citealt{belob06}). 
Therefore they must be as close to the 
Northern Arm - which is passing by the IRS~16 cluster from behind -
as the IRS~16 cluster is with respect to the mini-cavity.
It is therefore likely that their stellar wind may have a similarly strong effect on
the Northern Arm as it may have  
towards the mini-cavity.

\subsubsection{Stellar probes of the wind from the direction of SgrA*}
\label{probes}

A particularly interesting feature in this context is feature X7. Despite 
the high velocity component to the northwest, the cometary or bow-shock-like shape of the 
feature is pointing approximately toward SgrA*. 
The wind from the direction of SgrA* could be responsible for this shape.
If the star is moving within an area with no significant amounts of ISM then any 
supersonic wind from the direction of SgrA* could easily produce such a shape. 
If the amount of the ISM is not negligible then the wind  
should have a much higher velocity than the observed velocity of the feature.
Taking into account the slight displacement
of the line of symmetry of X7 from the line that connects it with SgrA*, a simple
calculation tells us that the velocity of the wind from SgrA* should be
at least one order of magnitude higher than the velocity of the feature.
In addition, approximately along the same line that connects SgrA* and X7, 
lies the feature X3 with a same type of a cometary appearance (see Fig.~\ref{X3}). 
The alignment of X3 and X7 with SgrA*, and the shape of both features, give a 
strong support to the central outflow hypothesis.        
Recent VLT NACO K$_S$-band polarization measurements \citep{meyer06b} have  
shown that source X7 is also polarized by 30$\pm$5\% along a position angle 
of -34$^o$$\pm$10$^o$.
Assuming that the polarization is due to Mie scattering of a non-spherical
dust distribution then an overall direction of the bow-shock's symmetry axis
is 56$^o$$\pm$10$^o$. This includes the direction towards SgrA*
(see the description of data acquisition and calibration in \citealt{eckart06}
 and \citealt{meyer06b}).

\section{Summary and Conclusions}
\label{summary}

We have presented L'-band (3.8 $\mu$m) adaptive optics observations of an
approximately 18$\times$18 arcsec$^2$ (0.7$\times$0.7 pc$^2$) region around SgrA* 
obtained with the NAOS/CONICA system at ESO VLT. 
The images reveal a large number of thin filaments associated with the mini-spiral, 
that could be due to the interaction of a fast wind with an ambient ISM.
The previous detection of some of the filaments in the radio regime, as well as in 
Br$\,$$\gamma$ and Pa$\,$$\alpha$ lines, shows that
the filaments are not only associated with the continuum emission by dust, but also 
with the ionized gas component of the ISM. 
High-resolution spectroscopy is needed that could show if there are shock-diagnostic lines
coming primarily from the filaments in order to confirm our interpretation. However, there 
are a number of observational challenges involved (e. g. extinction, stellar contamination, resolution etc.) 
that may preclude a clear answer at this time. 

We performed proper motion measurements of these features on the time basis of 
approximately 4 years. 
The filaments located west from SgrA* are curved with their convex sides westwards and 
 show a significant proper motion to the west. 
The comparison with the Keplerian orbit fitting of the Northern Arm presented in \citet{paum04},  
tells us that the process that
drives the motion of the filaments cannot be purely due to the gravitational potential
of the supermassive black hole at the center.

We propose the following scenario:
An at least partially collimated outflow is emanating from 
a combination from the disk of high mass-losing He-stars and an outflow associated 
with the black hole at the position of SgrA*.   
This scenario represents a unified model that would explain the mini-cavity, the two cometary shaped sources X3 and X7, 
as well as the northern and southern lobes in the NIR $H_2$ emission (see appendix \ref{outflow}).
It would also be in agreement with the recent NIR polarization data that is indicative 
for a temporal accretion disk around SgrA* and would be supportive for a
 collimated outflow that is often invoked to explain the
properties of SgrA*.

\appendix

\section{L'- and K$_S$-band stellar proper motion comparison}
\label{stellar pm}
\begin{table*}
\caption{Comparison of stellar proper motions in K$_S$- and L'-band. All velocities are in km$\,$s$^{-1}$.  } 
\begin{center}
\begin{minipage}{1.0\textwidth}
\begin{tabular}{lccc|cccccc|cccccc}
\hline\hline
&&&&$K_S$&&$L'$&&$G00$ \footnote{data from \citet{genzel00} \label{fn2}}&&$K_S$&&$L'$&&$G00$ $\footref{fn2}$\\
$name$ &r\footnote{relative to SgrA*; position in the 2003.356 epoch \label{fn1}} &  $\Delta \alpha$ \footref{fn1}& $\Delta \delta$ \footref{fn1} &
$v_{\alpha}$ & $\Delta v_{\alpha}$& $v_{\alpha}$ & $\Delta v_{\alpha}$ & $v_{\alpha}$& $\Delta v_{\alpha}$
 & $v_{\delta}$ & 
$\Delta v_{\delta}$ &  $v_{\delta}$ & $\Delta v_{\delta}$&$v_{\delta}$& $\Delta v_{\delta}$ \\
 & (arcsec)&(arcsec) &(arcsec) &&&&&&&&&&&&\\
\hline
         & 1.05 &  0.80 & -0.68  &   426 & 14 &    431 &  31 &  410 &  80 &   23 &  12 &     9 & 11 &   50 & 100\\
IRS 16C  & 1.21 &  1.10 &  0.50  &  -285 &  9 &   -336 &  22 & -330 &  39 &  268 &  11 &   222 &  6 &  353 &  34\\
         & 1.45 & -1.35 &  0.52  &   -52 &  9 &    -53 &  37 &  -58 &  95 & -181 &  10 &  -212 & 29 & -116 &  62\\
         & 2.07 &  1.90 & -0.82  &   228 & 14 &    218 &  38 &  154 &  84 &  100 &  13 &    62 & 24 &   13 &  44\\
IRS 16CC & 2.07 &  1.99 &  0.55  &   -48 &  9 &    -55 &  22 &  -58 &  35 &  183 &  10 &   174 &  7 &  234 &  27\\
         & 2.18 &  1.85 & -1.15  &   254 &  9 &    248 &  23 &  255 &  28 &   20 &  10 &    47 &  8 &  113 &  23\\
IRS 33N  & 2.20 & -0.05 & -2.20  &   103 &  9 &    131 &  31 &   70 &  48 & -309 &  10 &  -339 & 14 & -243 &  28\\
         & 2.66 & -2.63 &  0.42  &  -326 & 14 &   -285 &  62 &      &     & -178 &  15 &  -149 & 44 &      &    \\
         & 3.17 &  2.94 & -1.19  &   139 & 10 &    121 &  27 &  147 &  51 &   87 &  11 &    85 & 15 &  111 &  55\\ 
         & 3.20 &  0.66 & -3.13  &   230 &  7 &    262 &  25 &	   &     & -154 &   9 &  -147 & 14 &      &    \\   
         & 3.22 &  3.22 & -0.01  &    75 & 10 &     53 &  27 &	   &     & -352 &  11 &  -372 & 23 &      &    \\
	 & 3.39 &  1.66 & -2.96  &     0 & 10 &    -27 &  31 &      &     &  58  &  11 &    52 &  6 &      &    \\
	 & 3.47 &  3.24 & -1.25  &   -11 & 17 &    -11 &  27 &	   &     & -178 &  13 &  -175 & 16 &      &    \\
	 & 3.84 & -1.26 &  3.63  &   204 & 11 &    206 &  46 &      &     &  148 &  12 &   124 & 18 &      &    \\
         & 4.15 & -3.34 &  2.46  &  -123 & 12 &   -102 &  39 &      &     &  402 &  12 &   405 & 16 &      &    \\
	 & 4.19 &  4.19 &  0.11  &   181 & 12 &    177 &  29 &      &     &  -13 &  11 &   -26 & 25 &      &    \\
	 & 4.55 &  2.95 &  3.46  &   237 & 10 &    247 &  25 &      &     &  -52 &  10 &   -46 &  5 &      &    \\
         & 4.99 &  2.20 &  4.48  &   248 & 12 &    248 &  22 &      &     & -120 &  12 &  -118 &  6 &      &    \\
 IRS 2L  & 5.95 & -3.68 & -4.67  &  -118 & 11 &    -89 &  22 &	   &     & -252 &  12 &  -288 & 14 &      &    \\
         & 6.30 &  2.85 & -5.62  &   229 & 12 &    239 &  32 &      &     &    4 &  11 &    26 & 30 &      &    \\ 
         & 6.62 &  0.94 & -6.55  &   -55 & 11 &    -57 &  39 &      &     &  -48 &  11 &   -59 & 18 &      &    \\
         & 6.90 &  5.09 &  4.66  &   102 & 18 &    108 &  37 &      &     & -154 &  15 &  -148 & 20 &      &    \\
IRS 1NE  & 7.26 &  7.02 &  1.84  &   209 &  9 &    187 &  18 &      &     &  -45 &  10 &   -12 &  7 &      &    \\ 
IRS 1SE  & 7.51 &  7.50 & -0.45  &   150 &  9 &    110 &  17 &      &     &  -80 &  10 &   -85 & 25 &      &    \\
         & 7.84 & -5.51 & -5.57  &   -32 & 10 &    -59 &  20 &      &     &  198 &  11 &   193 & 21 &      &    \\
IRS 30E  & 8.00 & -5.69 &  5.62  &    87 & 10 &    117 &  31 &   60 & 60  &  149 &  10 &   105 & 19 &   75 &  30\\
	 & 8.30 &  3.95 & -7.30  &    98 & 10 &    108 &  32 &      &     & -374 &  11 &  -337 & 30 &      &    \\

\end{tabular}
\label{tabcomparison}
\end{minipage}
\end{center}
\end{table*}

In Section~\ref{methods} we explain how L'-band image frames from different epochs have been transformed to
the coordinate system of a reference frame using IDL image transformation routines. 
In order to validate our method and to demonstrate that $(i)$ reliable proper motions can be
obtained from the lower resolution L'-band data and $(ii)$ 
that the resulting transformed images therefore are indeed astrometric and suitable 
for the measurements conducted in this work,
we compare K$_S$-band and L'-band stellar proper motions.
The PSF fitting was done using the StarFinder code \citep{diolaiti00}. 
While the L'-band positions were obtained from transformed images
we derived the K$_S$-band positions from the original image frames and applied 
the corresponding transformation to the resulting lists.
So the basic difference between two sets of proper motions is in the transformation of 
data to the reference coordinate system.

\begin{figure}
  \resizebox{8.5cm}{!}{{\includegraphics{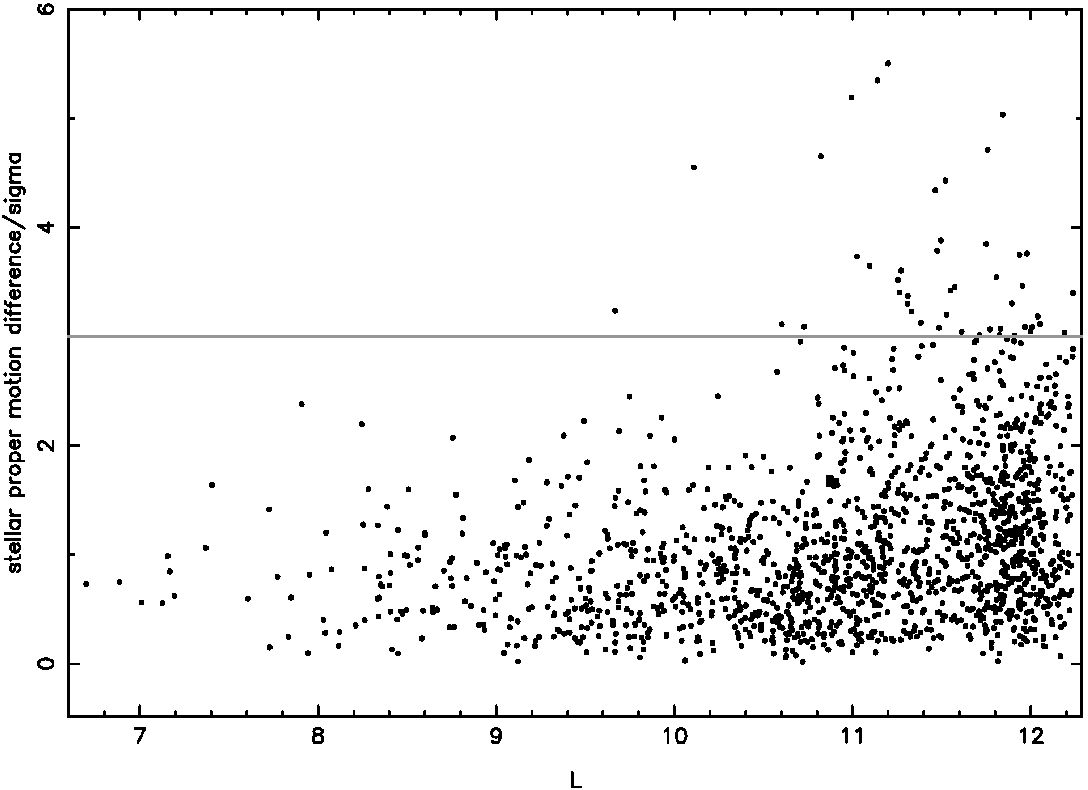}}}
  \caption{L'- and K$_S$-band stellar proper motion difference vs. L' magnitude. 
The y-axis labels are calculated via 
$\vert$$\vec{v_K}$-$\vec{v_L}$$\vert$/~$\sigma$($\vert$$\vec{v_K}$-$\vec{v_L}$$\vert$), where
 $\vec{v_K}$ and $\vec{v_L}$ represent the proper motion of a star as measured from K$_S$- and L'-band frames, respectively, and
 $\sigma$($\vert$$\vec{v_K}$-$\vec{v_L}$$\vert$) represents the standard deviation of the difference between two measurements.
 We find that 97~$\%$ of all the sources lie below the 3~$\sigma$ line.}
\label{comparison}
\end{figure}

\begin{figure}
  \resizebox{8.5cm}{!}{{\includegraphics{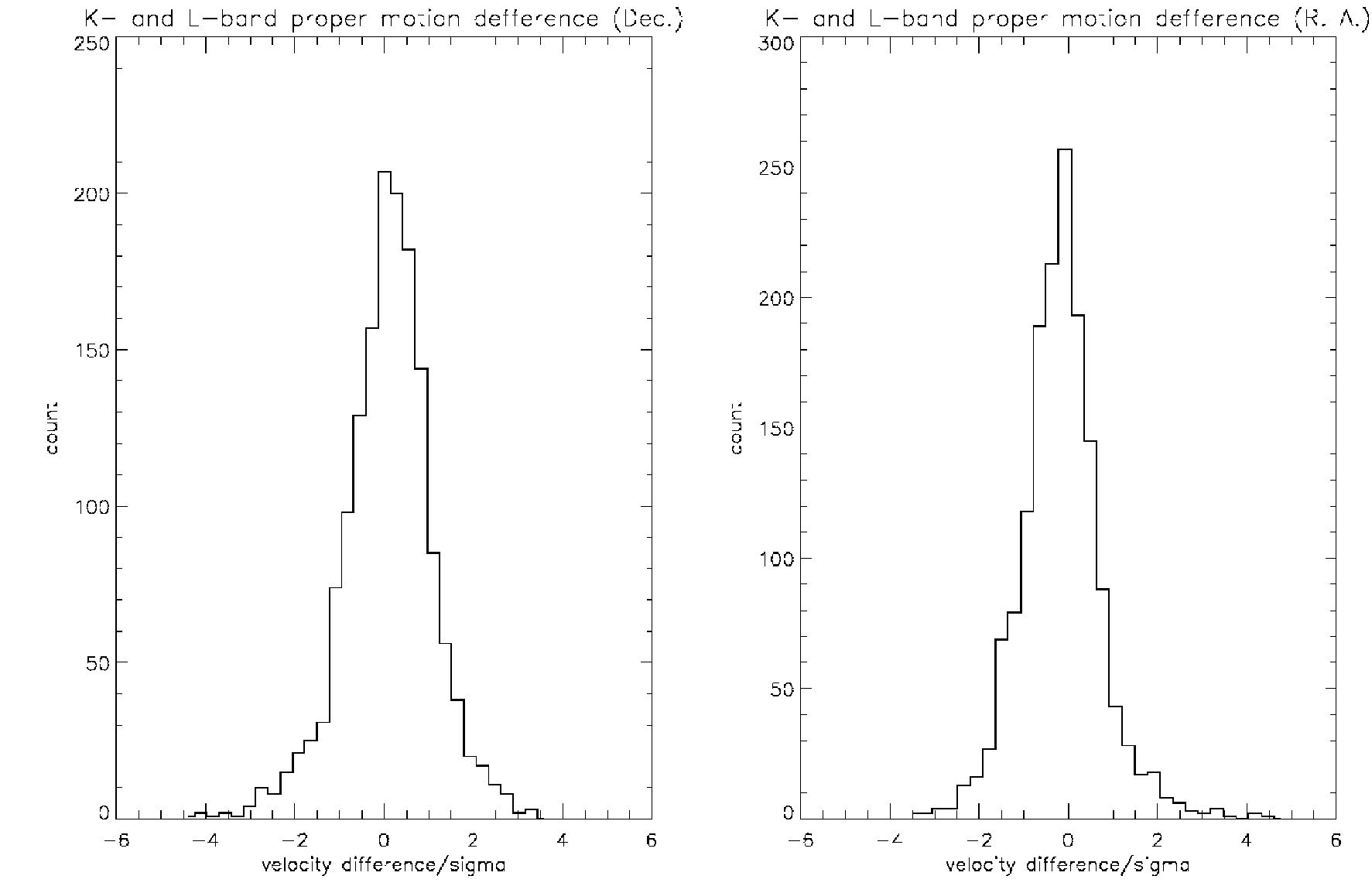}}}
  \caption{K$_S$- and L'-band proper motion difference projected in R. A. and Dec.}
\label{histogram}
\end{figure}

As can be seen from Fig.~\ref{comparison}, 97$\%$ of our K$_S$- and L'-band stellar proper motion 
results are in agreement to within 3$\sigma$. Also, there is no systematic offset between
proper motion difference when projected in R. A. and Dec. (see histograms in Fig.~\ref{histogram}).
 This demonstrates that the L'-band image transformation we used produces astrometric frames. 
As an example, we list in Table~A.1. proper motions for some of the sources in the central parsec obtained in our study 
(K$_S$- and L'). We also add the values published in \citet{genzel00}. The uncertainties of the proper motions 
represent the 1$\sigma$ deviation of the linear fit.

\section{Proper motion of the filament NE4}
\label{NE4}
\begin{figure}
  \resizebox{8.5cm}{!}{{\includegraphics{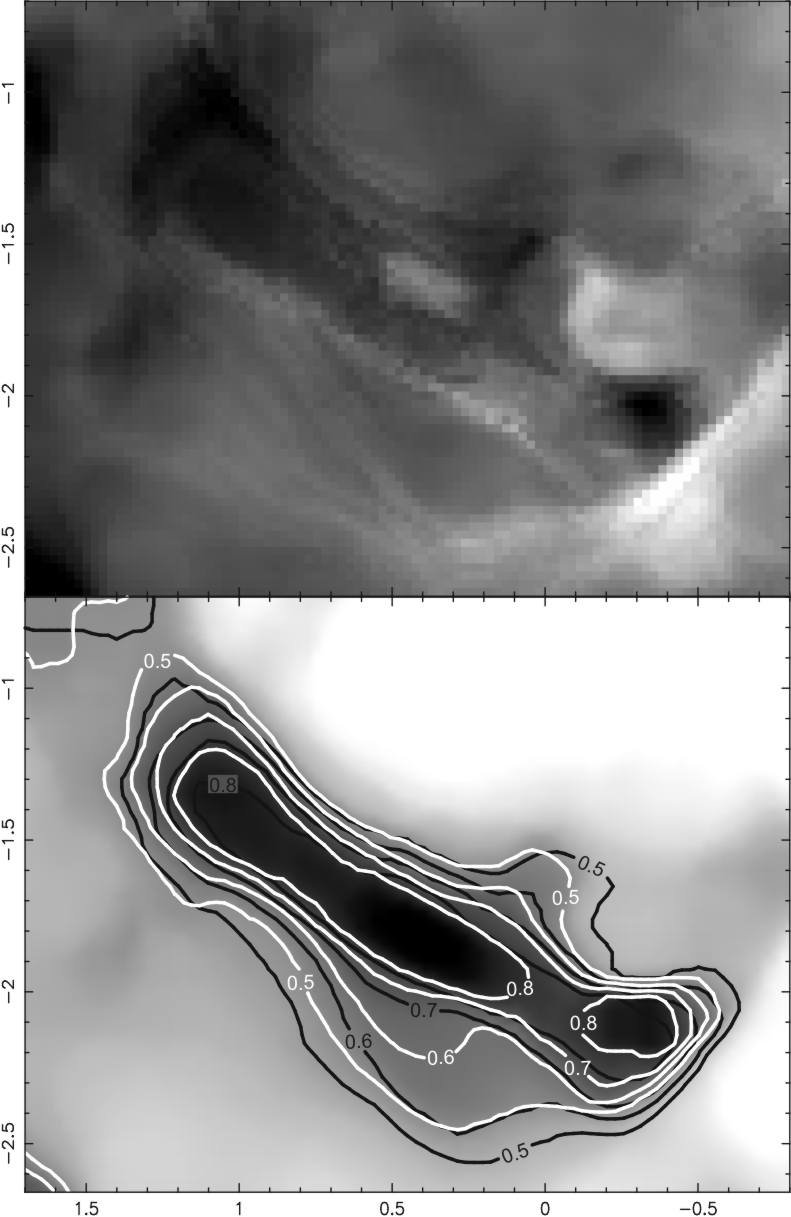}}}
  \caption{Upper panel: difference image of the area around the filament NE4,
 between the epochs 2006.408 and 2003.356. White pixels correspond to
 the later epoch, thus indicating the south-westward motion of the filament.
 Lower panel: grayscale image of the same area (2003.356) with the contours corresponding
to the 2006.408 (black) and 2003.356 epoch (white) overlayed. Contour levels correspond to 
the 50, 60, 70 and 80~\% of the maximum intensity of the image (both images were previously scaled to
have the same maximum intensity).}
\label{NE4fig}
\end{figure}
 The proper motion of the extended filament NE4 was derived 
 from images of the fitted background flux density which are free of 
 contaminating flux contributions from stars.
 Here we show, that the motion towards the south west can already
 be seen but is hard to quantify from the difference image (Fig. \ref{NE4fig}).
 From a comparison of contour plots the motion can be estimated
 to first order (Fig. \ref{NE4fig}). The actual value of the proper motion 
 was derived using the cross-correlation method described in the
 main text.

\section{A partially collimated outflow?}
\label{outflow}
\begin{figure}
  \resizebox{8cm}{!}{{\includegraphics{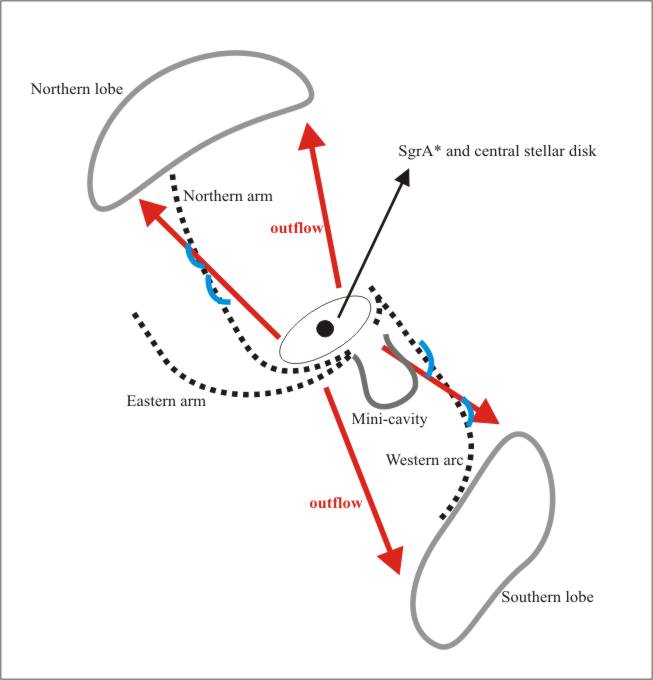}}}
  \caption{Sketch of the outflow model. Collimated outflow originating from the mass-losing stars cluster and possibly the black hole at the position of SgrA* could, in the interaction with the material of the mini-spiral, explain the geometry and motions of the thin filaments. This model suggests that northern and southern lobes of the CND are associated with the outflow.}
\label{model}
\end{figure}

\subsection{H$_2$ emission from the CND lobes}

\citet{gatley84, gatley86,burton92} and \citet{y-z01} 
report H$_2$(1-0) S(1) molecular hydrogen line emission from the CND that is
concentrated on the northern and southern lobes.
Based on HST NICMOS and ground based data and adopting a lower limit 
to the extinction toward the lobes of A$_K$=3 \citep{scoville03}, 
\citet{y-z01} derive the extinction-corrected peak line 
intensity from the lobes as approximately 
to be about a factor $\eta_1$=3-4  
higher than the mean surface brightness of the CND. 
The source of excitation of the H$_2$ emission associated with the 
CND is unclear.
Most commonly it is assumed that the $\sim$70$^o$ inclination 
of the 0.5~pc thick CND torus toward the line of sight will 
produce the north and south lobes by  
limb brightening of the $\sim$0.5~pc torus
\citep[e.g.][]{gatley86,burton92,y-z01}.
As summarized by \citet{y-z01}, neither the
intense UV filed combined with the high densities, nor the shock processes
 due to isotropic wind originating from the hot stars at the central parsec, can 
explain such a high intensity of molecular hydrogen emission on the CND lobes.
The latter consideration shows that the ram 
pressure of an isotropic wind from IRS~16 cluster would be at least 
a factor $\eta_2$=5 too low compared to what is required.

\subsection{Model}

In this section we consider the possibility that the outflow originating 
from the central part of 
the Galaxy, presumably from the cluster of high mass losing stars, with a 
possible contribution from the massive black hole itself,
 might actually be partially collimated.
In the case of the existence of a wind from SgrA*, the collimation would
 have to be seen as a consequence of the
yet unexplained accretion process onto the black hole but could be 
linked to the plane in which the mass-losing He-stars are moving in.
This motion within a single plane
implies that the viscous friction of gas being blown into that plane is
larger than that for gas being blown out perpendicular to it.
On the one hand that results in a pressure gradient perpendicular 
to the plane, on the other hand it will 
result in an accretion flow within that plane
onto the central black hole and may give rise to a corresponding 
outflow perpendicular to it. 
So both the ordered motion of the mass-losing stars as well as
the resulting structured accretion flow may result in a combined, 
possibly collimated outflow.

This scenario would also explain the bow shock ''tail'' of the IRS~7
and the northern and southern lobes 
within the CND to be associated with the outflow.
If we assume that the H$_2$ line emission from the lobes is dominated
by the interaction with a collimated wind rather than an approximately 
isotropic stellar wind from the central few arcseconds
then - due to the limited size of the two lobes in the CND - 
a geometrical concentration of the outflow over an isotropic flow by
a factor $\alpha$$\sim$50 has to be taken into account.
We note that such an outflow could also have a grazing incident onto the CND
(see Fig.~\ref{sphere}).
In this case the opening angle could be larger and the compression factor smaller.
We also infer the lower limit on the collimated outflow velocity from the fact that
NE1, the northern most filament, has an insignificant proper motion 
perpendicular to its extent and that filaments
NE3 and NE4 which are likely to be more affected by the IRS~16 cluster wind have
proper motion velocities of $v_{pm}$$\sim$150$\,$km$\,$s$^{-1}$.
When generated in the Northern Arm at angular distances of 3 to 7~arcseconds
from the IRS~16 cluster or SgrA* and taking $v_{pm}$ as an upper limit of the 
proper motion inferred by a collimated outflow, then a lower limit to the velocity of that flow 
with a total opening angle of $\alpha_{open}$=20$^o$-30$^o$ may be of the order of
$v_{f}$$\sim$$v_{pm}$/tan($\alpha_{open}/2$)$\sim$$\,$500-900$\,$km$\,$s$^{-1}$.
The total required mass-loss carried by the collimated outflow is then
\begin{equation}
\dot{m}_{flow} = \frac{\eta_1 \eta_2}{\alpha} \frac{v_s^2}{v_f^2} \dot{m}_{stellar}
\le 10^{-3} M_{\odot} yr^{-1}~~~.
\end{equation}  
This implies that such an outflow can easily be driven from the 
mass loss wind of the stellar cluster, that is blown away from the stellar disk
or SgrA* itself, rather than being accreted onto SgrA*.

Whether in the presence of the deep central potential 
provided by the 3.6$\times$10$^6$\solm black hole
the wind contribution is dominated by the stars or by SgrA* 
has to be investigated via detailed hydro-dynamical model calculations.
In the context of the mini-cavity such calculations have been 
carried out by \citet{melia96} and most recently by \citet{cuadra06}. 
The authors have shown that the structure, kinematics, and luminosity 
of the mini-cavity are consistent with 
a model in which a few 10$^6$\solm black hole 
interacts with an
ambient 500-700 km$\,$s$^{-1}$ Galactic Center wind originating within 
the IRS~16 cluster. 
It accretes a small portion of it and then may expel the rest 
in form of a collimated flow toward the mini-spiral streamer
\citep{melia96}.
The filaments along the Northern Arm presented
here, the shape of the Northern Arm, as well as the approximate alignments
between the rotation axis of the He-star disk and the connecting 
line between the northern and southern lobes in the CND 
(see references given above) present evidence for a similar collimated 
flow to the northeast.

\subsection{3-Dimensional Orientation}
\label{orientation}

\begin{figure}
  \resizebox{8cm}{!}{{\includegraphics{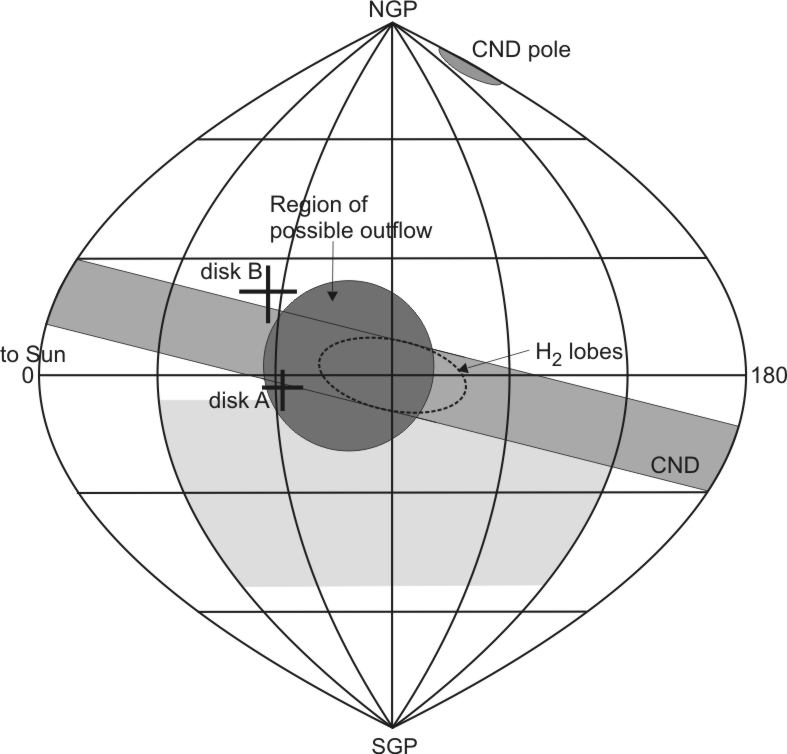}}}
  \caption{Positions of main features at the Galactic Center as seen by an observer at the 
center of the Galaxy. Two crosses indicate the normal to the clockwise stellar 
disk plane: disk~A as found by \citet{l&b03} and disk~B by  
\citet{paum06}. The large light-gray-shaded area south of the galactic plane 
represents the results from NIR polarization measurements as presented by 
\citet{eckart06} and \citet{meyer06a, meyer06b}. Here we only show the direction
 towards the Northern Arm.
The dark-gray-shaded area close to the center of the plot outlines the approximate 
direction of a possible collimated outflow that could explain out new observations
plus the H$_2$ results by 
\citet{gatley84, gatley86, burton92} and \citet{y-z01}. 
Such an outflow would also be in agreement with mass-loss from the the stellar disks 
and the recent polarization data (see references above).
}    
\label{sphere}
\end{figure}

In the following we discuss the space orientation of the main structures at
the Galactic Center, as well as the orientation of the proposed outflow.
  
The observed geometry in the plane of the sky is such that the Northern Arm filaments and the 
mini-cavity are on almost opposite sides and on a well defined 
connection line through SgrA*. This line also coincides with the strip of low extinction 
presented by \citet{schoedel07}. The wind that we consider responsible for 
the filaments along the Northern Arm and 
 the formation of the mini-cavity as well, 
is strongly interacting with the gas in the gas disk $(i)$
presented by \citet{vol&dus00}. This disk contains the 
Northern Arm and the part of the Bar, and connects well with the CND at an inclination
 of $\sim$30$^o$ to the line of sight. The disk $(i)$ is inclined $\sim$25$^o$ to 
the line of sight and has 
the position angle of 28$^o$ on the sky (N to E). 
Both CND and the Northern Arm are at a 
relatively low inclination 
to the plane of the Galaxy \citep{vol&dus00,paum04}. 
Thus we conclude that the wind could in fact be 
collimated with an opening angle of the order of 30$^o$ or less. 
Another indication for an outflow to the northeast
comes from the wavy structure of the Northern Arm.
Here \citet{y-z93} already noted that
the distortions of the arm are produced by ram 
pressure of the wind incident on the surface of 
the arm and inducing there Rayleigh-Taylor or 
Kelvin-Helmholtz instabilities.

The position angle of the outflow projected on the sky is close to that 
of the galactic plane. This would also be in good agreement with the extent of the 
northern and southern lobes in the H$_2$ emission of the CND. 
These lobes are located at the outer tips of the projected ellipse of the 
rotating CND. Therefore the lobes must be located approximately
within the plane of the sky and within the Galactic plane.

This comparison also shows that the rotation axis of the hot He-stars
\citep{genzel00,genzel03,l&b03,paum06}
and the axis of the outflow may in fact be aligned to within less then 30$^o$.
Both results for the disk of clockwise-moving stars from  \citealt{l&b03} (disk~A)
 and \citealt{paum06} (disk~B) are shown in the Fig.~\ref{sphere}. The rotational axis of 
the disk~A is closer to the direction of the postulated outflow than the axis of the 
disk~B. The reason for this behavior could be the choice of the stellar sample in both
 measurements. 
The sample in \citet{l&b03} includes 10 of the 13 brightest 
brightest He stars, most of them designated as Ofpe/WN9 stars by \citet{paum06}. 
Disk~A therefore contains the largest number of heavily mass-loosing stars.

\citet{eckart06} obtained polarization measurements of SgrA* at two observing epochs 
(July 2005 and June 2004) and report the mean position angle of the E-vector to be about 
60$^o$ with a swing in polarization angle of up to 40-50$^o$. 
\citet{meyer06a} find that the inclination of the possible accretion disk around SgrA* has 
a lower limit of about 30$^o$. 
It is interesting to point out
that the range over which polarization angle varies on the plane of the sky, centered on SgrA*
(see Fig.~8 in \citealt{eckart06}), approximately  coincides with the opening angle of the central
outflow proposed in this work. These two phenomena could be connected under the assumption that the accretion disk,
 or the possible jet, may be polarized and the
E-vector is positioned perpendicularly to the accretion disk, or along the jet 
(see the discussion in \citealt{eckart06} and \citealt{meyer06a}).   
Fig~\ref{sphere} shows the orientations of the discussed structures 
for an observer at the center of the Galaxy. 

\acknowledgements{We thank P. Pejovic for providing his K-band stellar proper motion measurements.
Part of this work was supported by the $Deutsche Forschungsgemeinschaft$ (DFG) via SFB 494. 
K. Mu\v{z}i\'{c} and L. Meyer
 were supported for this research through a stipend from the International Max Planck Research School
(IMPRS) for Radio and Infrared Astronomy at the Universities of Bonn and Cologne.}
\bibliography{6265man}

\end{document}